\documentclass[aps,prl,twocolumn,floats,epsfig,pdflatex]{revtex4}
\usepackage{epsfig}
\usepackage{amsmath}
\usepackage{amsfonts}
\usepackage{graphicx}
\usepackage{amssymb}
\usepackage{amsbsy}
\usepackage{xcolor}
\usepackage{ulem}

\newcommand{\bra}[1]{\langle{#1}|}
\newcommand{\ket}[1]{|{#1}\rangle}

\newcommand{\beq}{\begin{equation}}
\newcommand{\eeq}{\end{equation}}
\newcommand{\bea}{\begin{eqnarray}}
\newcommand{\eea}{\end{eqnarray}}

\newcommand{\beginsupplement}{%
        \setcounter{table}{0}
        \renewcommand{\thetable}{S\arabic{table}}%
        \setcounter{figure}{0}
        \renewcommand{\thefigure}{S\arabic{figure}}%
        \setcounter{equation}{0}
        \renewcommand\theequation{S\arabic{equation}}%
     }
\begin{document}

\title{Collapse and revival of quantum scars via Floquet engineering}
\author{Bhaskar Mukherjee$^1$, Sourav Nandy$^1$, Arnab Sen$^1$, Diptiman
Sen$^2$, and K. Sengupta$^1$}
\affiliation{$^1$School of Physical Sciences, Indian Association for the
Cultivation of Science, Kolkata 700032, India \\
$^2$Centre for High Energy Physics and Department of Physics, Indian
Institute of Science, Bengaluru 560012, India}
\date{\today}

\begin{abstract}
The presence of quantum scars, athermal eigenstates of a many-body
Hamiltonian with finite energy density, leads to absence of
ergodicity and long-time coherent dynamics in closed quantum systems
starting from simple initial states. Such non-ergodic coherent
dynamics, where the system does not explore its entire phase space,
has been experimentally observed in a chain of ultracold Rydberg
atoms. We show, via study of a periodically driven Rydberg chain,
that the drive frequency acts as a tuning parameter for several
reentrant transitions between ergodic and non-ergodic regimes. The
former regime shows rapid thermalization of correlation functions
and absence of scars in the spectrum of the system's Floquet
Hamiltonian. The latter regime, in contrast, has scars in its
Floquet spectrum which control the long-time coherent dynamics of
correlation functions. Our results open a new possibility of drive
frequency-induced tuning between ergodic and non-ergodic dynamics in
experimentally realizable disorder-free quantum many-body systems.
\end{abstract}


\maketitle

The eigenstate thermalization hypothesis (ETH) is one of the central
paradigms for understanding out-of-equilibrium dynamics of closed
non-integrable quantum systems \cite{rev1a,rev1b,rev1c,rev1d,
rev2,deutsch1,srednicki1,rigol1}. It posits that {\it all} bulk
eigenstates of a generic quantum many-body Hamiltonian are thermal;
their presence ensures ergodicity and leads to eventual
thermalization for out-of-equilibrium dynamics of a generic
many-body state \cite{rev2}. This hypothesis is strongly violated in
certain cases, the most famous example being one-dimensional (1D)
disordered electrons in their many-body localized phase
\cite{mblref1,mblref2}. More recently another example of a weaker
failure of ETH, due to the presence of quantum many-body scar
states, has been studied extensively in disorder-free systems
\cite{scarrefqm1,scarref1,scarref2a,scarref2b,scarref2c,scarref2d,scarref2e,
scarref3a,scarref3b,scarref3c,scarref3d,scarref3e}. Scars are
eigenstates with finite energy density but anomalously low
entanglement \cite{scarref1,scarref2b,scarref2c,scarref2d,scarref3b}
which form an almost closed subspace in the system's Hilbert space
under the action of its Hamiltonian. Their presence leads to
persistent coherent oscillatory dynamics of correlation functions
starting from initial states that have a high overlap with scars.
This provides an observable consequence of their presence as
verified in recent experiments on quench dynamics of a chain of
ultracold Rydberg atoms \cite{scarref1}.

Here we study the fate of such ergodicity violation in a
periodically driven Rydberg chain. It is well known that the
stroboscopic dynamics of a periodically driven quantum system is
controlled by its Floquet Hamiltonian $H_F$ \cite{rev3} which is related to its
unitary evolution operator $U$ through $U(T,0) = \exp[-i H_F
T/\hbar]$, where $T=2 \pi/\omega_D$ is the time period of the drive
and $\omega_D$ is the drive frequency. For generic disorder-free
systems, such driving is expected to cause thermalization to a
featureless ``infinite temperature'' state \cite{steady1b,steady1c,steady2,
steady3}.
In what follows, we
will study the possibility of the existence of scars in the eigenstates
of $H_F$ as a function of $\omega_D$ and relate their influence on
the dynamics of correlation functions. Our initial state will be an
experimentally realized ${\mathbb Z}_2$ symmetry broken many-body
state which has one Rydberg excitation in alternate lattice sites
\cite{scarref1,rydramp1,rydramp2,rydramp3}.

The central results of this study are as follows. First, for large
$\omega_D$ and starting from a initial $\mathbb{Z}_2$ state, we show
the presence of long-time persistent oscillations of the density-density
correlator of Rydberg atoms. Such oscillations have characteristic
frequencies which are different from $\omega_D$ indicating a lack of
synchronization (a hallmark of thermalization in periodically driven
systems). We relate this oscillation frequency to the quasienergy separation
between the scar states of the Floquet Hamiltonian indicating the
central role of these states in the dynamics. Second, at ultra-low
drive frequencies, we find that there are no persistent oscillations,
and the behavior of the correlator agrees with that expected from
ETH. We show numerically that in this regime, there are no scars in
the eigenspectrum of $H_F$ and the dynamics is controlled by a set
of thermal states. Finally, we find several drive-frequency-induced
transitions between thermal and coherent regimes at intermediate frequencies.
These transitions, that have no analogs in the non-driven systems studied
earlier~\cite{scarref1,scarref2a,scarref2b,scarref2c,scarref2d,scarref2e,
scarref3a,scarref3b,scarref3c,scarref3d,scarref3e}, provide a route
to controlled switching between ergodic and non-ergodic dynamics of
the Rydberg atoms. We chart out the critical drive frequencies at
which these transitions occur, provide an analytic understanding of
their occurrence, and suggest experiments which can test our theory.

{\it Model}: The low-energy properties of an ultracold Rydberg atom
chain can be described by an effective two-state Hamiltonian on each
site given by \cite{scarref1,rydramp1,rydramp2,rydramp3}
\begin{eqnarray} H_{\rm RYD} &=& \sum_i ~(\Omega \sigma_i^x + \Delta n^r_i) ~+~
\sum_{ij} ~V_{ij} n^r_i n^r_j. \label{rydham1} \end{eqnarray}
The two states correspond to the ground ($|g_i\rangle$) and Rydberg
excited states ($|e_i\rangle$) of the atoms on site $i$. The dipole
blockade in these systems ensures that there is at most one Rydberg
excitation per site: $n_r^i \le 1$, where $n^r_i = (1 + \sigma^z_i)/2$
is the Rydberg excitation number operator, and $\sigma_i^x =
|g_i\rangle \langle e_i|+ |e_i\rangle \langle g_i|$ denotes a Pauli
matrix on site $i$ which couples the ground and excited states. In
Eq.\ \eqref{rydham1}, $\Delta$ is the detuning parameter which can be
used to excite an atom to its Rydberg state, $V_{ij}$ denotes
an interaction between two Rydberg excitations, and $\Omega$ is the
coupling strength between ground and excited states. In experiments,
it is possible to reach a regime where $V_{i,i+1} \gg \Omega, \Delta \gg
V_{i,i+2}$ so that the Hamiltonian of the model
becomes~\cite{scarref1,dipoleexp1,dipoleexp2,subir1,subir2}
\begin{eqnarray} H'_{\rm RYD} &=& \sum_i ~(\Omega \sigma_i^x + \Delta n^r_i),
\label{rydham2} \end{eqnarray}
and this is to be supplemented by the constraint that $n^r_i n^r_{i+1}=0$
for all sites $i$.

This constrained model can be easily mapped into an Ising-like spin
model in the presence of both longitudinal and transverse fields of
strength $\Delta$ and $\Omega$ respectively. Within the constrained
Hilbert space of the system, one can represent $H_{\rm RYD}$ as
\cite{scarref2c,scarref2d}
\begin{eqnarray} H_{\rm spin} &=& \sum_{i} \left(- w \tilde \sigma_{i}^x +
\frac{\lambda}{2} \sigma_{i}^z \right), \label{spinham}
\end{eqnarray} where $P_i=(1-\sigma_i^z)/2$ is a local projection
operator, ${\tilde \sigma}^{\alpha}_i = P_{i-1} \sigma_i^{\alpha}
P_{i+1}$ and $\alpha=x,y,z$, $\Omega \equiv -w$, and $\lambda \equiv
\Delta$. Our analysis will be based on this model. Eq.\
\eqref{spinham} also provides a low-energy description for the 1D
tilted Bose-Hubbard model as detailed in the Supplementary
Information. For $\lambda=0$, $H_{\rm spin}$ reduces to the ``PXP
model'' studied in Refs.\ \cite{scarref2c,scarref2d} which is known
to host quantum scars among its eigenstates.

{\it Analysis}: We analyze the periodic dynamics of $H_{\rm spin}$
for a square pulse protocol: $\lambda(t) = -(+) \lambda$ for $t\le
(>)T/2$. The unitary evolution operator at the end of a drive cycle
can be written as $U(T,0) = e^{-i H_{\rm spin} [\lambda]T/2} e^{-i
H_{\rm spin}[-\lambda] T/2}$. The evolution operator can then be
expressed as
\begin{eqnarray} U(T,0) &=& \sum_{p,q} e^{-i(\epsilon_q^+ + \epsilon_p^-)T/2}
c_{pq}^{-+} |p^-\rangle \langle q^+| \label{uevol1}, \end{eqnarray}
where $\epsilon_p^{+(-)}$ and $|p^{+(-)}\rangle$ are eigenstates and
eigenfunctions of $H_{\rm spin} [+(-) \lambda]$ and $c_{pq}^{-+} =
\langle p^-|q^+\rangle$. The spin correlation function
$O_{ij}=\langle (1+\sigma^z_{i})( 1+\sigma^z_{i+j})\rangle/4$ of the
spins between any two sites $i$ and $i+j$ can then be obtained,
after $n$ drive cycles, as
\begin{eqnarray} O_{ij} &=& \sum_{p ,q} e^{-i n(\epsilon_p^- -\epsilon_q^+)T/2}
(c_{\psi_0 p}^{- \ast} c_{q \psi_0}^+)^n \langle p^-|O_{ij}|q^+\rangle,
\label{correxp1} \end{eqnarray}
where $c_{\psi_0 p}^{a} = \langle \psi_0 |p^a\rangle$ for $a=\pm$,
and $|\psi_0\rangle$ is the initial state. Unless explicitly stated
otherwise, we will choose $|\psi_0\rangle = |\mathbb{Z}_2\rangle=
| \cdots \downarrow \uparrow \downarrow \uparrow \cdots \rangle$ to be a
$\mathbb{Z}_2$ symmetry broken state, with $\langle \psi_0
|\sigma^z_{j} |\psi_0 \rangle = (-1)^{j+1}$. We note that $O_{ij}$
provides direct information of the density-density correlation
function of the Rydberg atoms after $n$ cycles of the drive. Our
numerical analysis will involve computation of $\epsilon_p^{\pm}$
and $|p^{\pm} \rangle$ using exact diagonalization for finite
chains of size $L \le 26$ and subsequent evaluation of $O_{ij}$
using Eq.\ \eqref{correxp1}.

To obtain an analytical understanding of the nature of the dynamics,
we derive the Floquet Hamiltonian corresponding to $U = \exp[-i H_F
T/\hbar]$ using a Magnus expansion which is expected to yield an
accurate description of the dynamics for high drive
frequencies~\cite{rev3}. Further details are provided in Supplementary 
Information. To $\mathcal{O}(1/\omega_D^3)$, this calculation
leads to $H_{F}^{\rm Magnus} = H_0 + H_1$, where
\begin{eqnarray} H_0 &=& -w \sum_{j} ~\left[ C_1 \tilde \sigma_{j}^x
+ C_2 \tilde \sigma_{j}^y \right], \label{magexp1} \\
H_1 &=& -\frac{2 \lambda \delta^3}{3} \sum_{j} ~\left[
\tilde{\sigma}_{j-1}^y {\tilde \sigma_{j}^{z}}+ {\tilde
\sigma_{j}^z} \tilde{\sigma}_{j+1}^y -\left(\tilde \sigma_{\ell}^y
\sigma_{\ell+1}^z P_{\ell+1} \right. \right.\nonumber\\
&&\left. \left. + \sigma^z_{\ell-1} P_{\ell-1} \tilde \sigma_{\ell}^y \right)+
\left(\tilde{\sigma}_{j}^y\tilde{\sigma}_{j+1}^y +\tilde{\sigma}_{j}^x
\tilde{\sigma}_{j+1}^x\right) \tilde{\sigma}_{j+1}^y /2 \right]. \nonumber
\end{eqnarray}
Here we have defined dimensionless quantities $\gamma= \lambda T/4
\hbar$ and $\delta = w T/(4 \hbar)$, $C_1 = 1- 2 \gamma^2/3$, and
$C_2= \gamma [1- (\gamma^2 - 4 \delta^2)/3]$. We find that the terms in
$H_0$ (which we denote as PXP terms) constitute a renormalized PXP
model (up to a global spin rotation); consequently, for $\hbar \omega_D
\gg \lambda, \delta$, where the effect of $H_1$ can be ignored, we
expect $H_F^{\rm Magnus}$ to host scar states similar to those in
the PXP model. However at moderate $\omega_D$, the terms in $H_1$
(which we denote as non-PXP terms), are expected to
become important. The competition between these two classes of terms
can be tuned using the drive frequency and will be discussed in detail below.

In what follows, we will be interested in large drive amplitudes for
which $\lambda \gg w$ ($\gamma \gg \delta$). In this regime, as
detailed in Supplementary Information, the Floquet Hamiltonian can be
perturbatively calculated to $\mathcal{O}(w)$ for an {\it arbitrary}
$\omega_D$ and gives {\it only} PXP terms with all non-PXP terms
(and further PXP terms) generated at $\mathcal{O}(w^2/\lambda)$ and
beyond. To $\mathcal{O}(w)$, the Floquet Hamiltonian equals
\begin{eqnarray} H_F ~=~ -w ~\frac{\sin \gamma}{\gamma} ~\sum_j ~[\cos
\gamma ~\tilde \sigma_{j}^x ~+~ \sin \gamma ~\tilde \sigma_{j}^y].
\label{highPXP} \end{eqnarray}
Eq.\ \eqref{highPXP} will be used to understand the transitions
between ergodic and non-ergodic regimes.

{\it Results}: To demonstrate the presence of ergodic to non-ergodic
transitions as a function of the drive frequency $\omega_D$, we
first compute the dynamics of the correlators starting from
$|\mathbb{Z}_2\rangle$. For this, we perform exact diagonalization
and compute $O_{22}$ from Eq.\ \eqref{correxp1} as a function of the
stroboscopic time $n$ (number of drive cycles) for several
$\omega_D$. In addition we also compute the half-chain entanglement
entropy $S_{L/2}$ for the eigenstates of the Floquet spectrum by
obtaining these via numerical diagonalization of $U$ (Eq.\
\eqref{uevol1}). This is followed by a computation of the reduced
density matrix for these eigenstates for a half chain from which
$S_{L/2}$ can be obtained using a standard procedure (see Supplementary 
Information). Quantum scars have $S \sim \ln L$ and are thus
expected to have much lower entanglement compared to thermal states
for which $S \sim L$. Thus $S_{L/2}$ provides a reliable way to
distinguish between thermal states and scars for a finite-size
many-body system.

\begin{figure}
\rotatebox{0}{\includegraphics*[width=\linewidth]{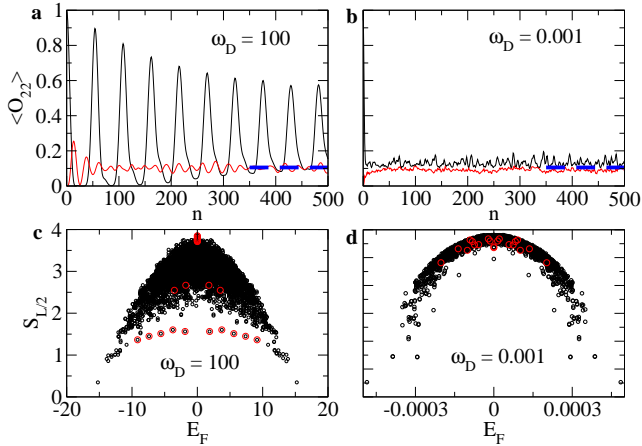}}
\caption{Plots of the correlator $O_{22}$, starting from an initial
state $|\mathbb{Z}_2\rangle$ (black solid lines), as a function of
the number of drive cycles $n$ for large (panel {\bf a}) and small
(panel {\bf b}) drive frequencies. The high-frequency regime shows
persistent long-time oscillatory dynamics while the low frequency
regime displays rapid thermalization consistent with the prediction
of ETH. The red solid lines correspond to plots of $O_{22}$ starting
from the initial state $|0\rangle = |\downarrow \downarrow
\downarrow ...\rangle$ and always display rapid thermalization
consistent with ETH. Panels {\bf c} and {\bf d}: Plots of half-chain
entanglement $S_{L/2}$ as a function of the Floquet quasienergy
$E_F$ for the Floquet eigenstates. The high [low] frequency regime
(panel {\bf c} [{\bf d}]) shows a clear presence [absence] of scars.
The red points correspond to eigenstates $|\psi_n\rangle$ which have
high ($|\langle \mathbb{Z}_2|\psi_n\rangle|^2 > 10^{-2}$) overlaps
with the initial state. All energies (frequencies) are scaled in
units of $w/\sqrt{2}$ ($w/(\hbar \sqrt{2})$), and we have chosen
$L=18$ and $\lambda=15$ in rescaled units for all plots.}
\label{fig1} \end{figure}

The results of these calculations are shown in Fig.\ \ref{fig1}.
Panel {\bf a} [{\bf b}] of Fig.\ \ref{fig1} shows the behavior of
$O_{22}$ as a function of $n$ for $\hbar \omega_D \gg [\ll] \lambda,
w$. We find that for $\hbar \omega_D \gg \lambda$, the dynamics
exhibits long-time coherent oscillations as expected from the quench
dynamics of the PXP model at $\lambda=0$ studied earlier
\cite{scarref2c,scarref2d}. The frequency of these oscillations
differs from $\omega_D$ indicating a clear lack of synchronization.
This behavior is expected from Eq.~\eqref{magexp1} where the non-PXP
terms appear in ${\mathcal O}(1/\omega_D^3)$ and hence are small.
The presence of scars in the Floquet Hamiltonian in this regime may
be confirmed from Fig.\ \ref{fig1} ({\bf c})which shows $S_{L/2}$
for Floquet eigenstates (denoted by $|\Phi_n\rangle$ henceforth) as
a function of the Floquet quasienergies $E_F$. The scar states are seen
as clear outliers in this plot. The eigenstates $|\Phi_n\rangle$
with large overlaps with $\mathbb{Z}_2$ ($|\langle
\mathbb{Z}_2|\Phi_n\rangle|^2 \ge 0.01$) are circled in red; from
this we find that the scars have maximal overlap with $|\mathbb
{Z}_2\rangle$ and thus control the dynamics leading to violation of
ETH \cite{scarref2d}.

In contrast, for $\hbar \omega_D /w \ll 1$, all the states including
those controlling the dynamics are thermal (Fig.\ \ref{fig1} ({\bf
d})). Consequently, there are no persistent oscillations for
$O_{22}$ (Fig.\ \ref{fig1} ({\bf b})) and one finds thermalization
consistent with ETH. We also note that the oscillatory behavior seen
in Fig.\ \ref{fig1} is a property of the initial
$|\mathbb{Z}_2\rangle$ state; a similar study of the dynamics for any
drive frequency starting from the Rydberg vacuum state $|0\rangle =
|\downarrow \downarrow \downarrow ...\rangle$ always provides fast
thermalization consistent with ETH (red curves in Fig.\
\ref{fig1} ({\bf a}) and ({\bf b})).

\begin{figure}
\rotatebox{0}{\includegraphics*[width=\linewidth]{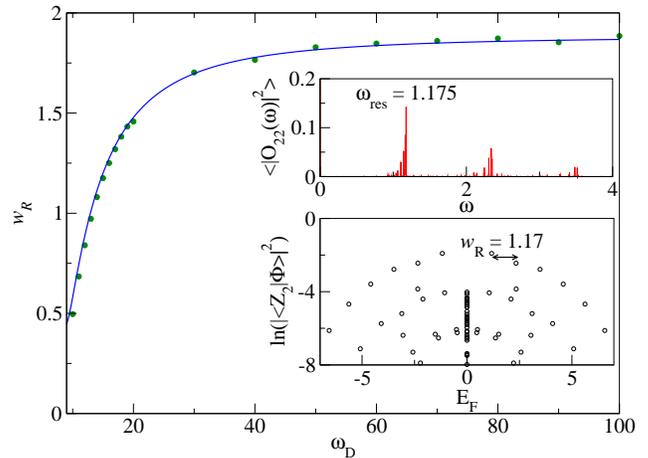}}
\caption{Plot of the scar quasienergy separation $w_R$ as a function
of $\omega_D.$ The green dots shows exact numerics while the blue
line shows the result of Eq.\ \ref{highPXP}. The top inset shows the
Fourier transform of the correlation function $|O_{22}(\omega)|^2$,
showing clear peaks at $\omega_{\rm res}$ and its multiples. The
bottom inset shows a plot of the overlaps of Floquet eigenstates
$|\Phi_n\rangle$ with the initial state ($|\langle
\mathbb{Z}_2|\Phi_n\rangle|^2$) as a function of $E_F$ displaying
the quasienergy separation ($w_R$), between the scar states which
have maximal overlaps with $|\mathbb{Z}_2\rangle$.
For both insets $\omega_D=15$. All energies
(frequencies) are scaled in units of $w/\sqrt{2}$ ($w/(\hbar
\sqrt{2})$), $L=18$ and $\lambda=15$ in rescaled units for all
plots.} \label{fig2}
\end{figure}

The quasienergy separation $w_R$ of the scars for $\hbar \omega_D \ge
\lambda$ as a function of $\omega_D$ is shown in Fig.\ \ref{fig2}.
We note that $w_R$ starts to decrease when $\gamma \to 1$; such a
behavior follows from the decrease of the norm of the PXP terms
in $H_F$ at $\mathcal{O}(w)$ with increasing $\gamma$ (Eq.\ \eqref{highPXP})
which gives $w_R = (\sin(\gamma) w_{\infty})/\gamma$, where
$w_{\infty}$ is the scar quasienergy separation for the undriven PXP model.
Fig.\ \ref{fig2} shows the exact match of the
numerical result with that obtained analytically.
The lowering of $w_R$ implies a sharp decrease in the oscillation
frequency $\omega_{\rm res}$ of $O_{22}$ as a function of $\omega_D$
in this regime. To check this, we extract $\omega_{\rm res}$ as a
function of $\omega_D$ from the Fourier transform of $O_{22}$ (upper
inset of Fig.\ \ref{fig2}) which matches the corresponding values of
$w_R/\hbar$ almost perfectly (lower inset of Fig.\ \ref{fig2}) and
shows a clear decrease with $\omega_D$. This provides a
drive-induced control over the quasienergy separation of the scars and
hence on the oscillation frequency which has no analog in earlier
quench studies. (Interestingly, the bottom inset of Fig.\ \ref{fig2} shows
a large number of states with zero quasienergy. They arise due to a symmetry
of the Floquet operator as discussed in Supplementary Information).

Next, we analyze the regime $\hbar \omega_D < \lambda$ where we
encounter the reentrant transitions between coherent and thermal
regimes. Here, we follow Ref.\ \cite{scarref2c,scarref2d} and use
the state $|\Psi_0 \rangle = (|\mathbb{Z}_2\rangle+
|{\bar{\mathbb{Z}}}_2 \rangle)/\sqrt{2}$ as our initial state, where
$|{\bar{\mathbb{Z}}}_2\rangle$ denotes the spin-flipped version of
$|\mathbb{Z}_2\rangle$. This allows us access to larger chain length
$L \le 26$ since $|\psi_0\rangle$ has weight only in the sector with
zero total momentum and spatial inversion (parity) symmetry.

\begin{figure}
\rotatebox{0}{\includegraphics*[width=\linewidth]{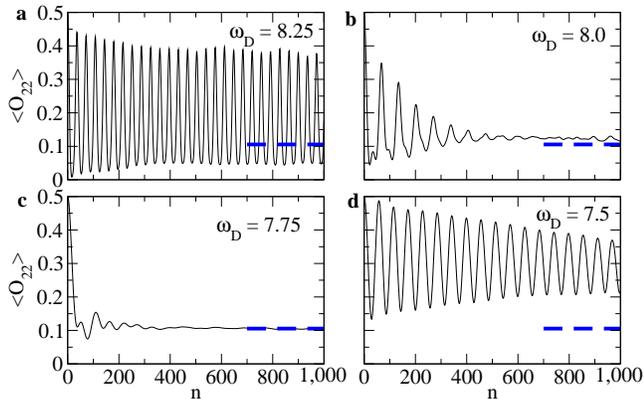}}
\caption{Plots of the correlator $O_{22}$ as a function of $n$ near
the transition at the largest drive frequency starting from an
initial state $|\psi_0\rangle= (|\mathbb{Z}_2\rangle+
|{\bar{\mathbb{Z}}}_2 \rangle)/\sqrt{2}$. The figure clearly
demonstrates a frequency-induced transition between non-ergodic and
ergodic regimes. The dynamics at $\omega_D=8.25$ (panel {\bf a}) and
$\omega_D=7.5$ (panel {\bf d}) shows persistent oscillations which
is inconsistent with the prediction of ETH and ergodic behavior. In
contrast, at $\omega_D=8.0$ (panel {\bf b}), these oscillations
dampen showing a precursor to ergodic behavior as predicted by ETH.
The dynamics at $\omega_D=7.75$ (panel {\bf c}) shows clear ergodic
behavior with fast thermalization time and almost no coherent
dynamics. The blue dashed line in all panels corresponds to the
infinite temperature value of $O_{22}$ as
predicted by ETH. For all plots $\lambda=15$, $L=26$ and all
energies (frequencies) are scaled in units of $w/\sqrt{2}
~(w/(\hbar\sqrt{2}))$.} \label{fig3} \end{figure}

\begin{figure}
\rotatebox{0}{\includegraphics*[width=\linewidth]{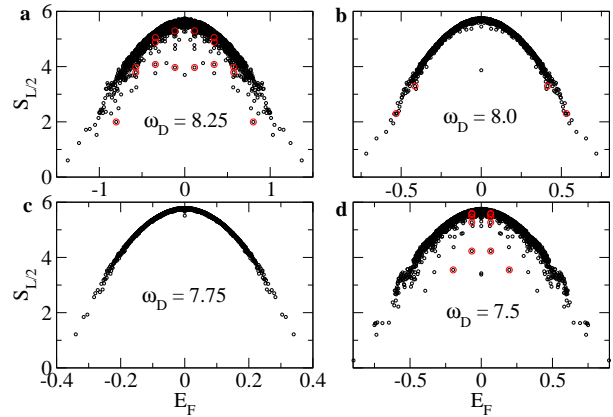}}
\caption{Plots of half-chain entanglement $S_{L/2}$ near the
transition at the largest drive frequency showing the presence and
absence of scars in the Floquet eigenspectrum. The entanglement at
$\omega_D=8.25$ (panel {\bf a}) and $\omega_D=7.5$ (panel {\bf d})
clearly indicate the presence of scars that have high overlaps with
the initial state ($|\langle \psi_0|\psi\rangle|^2 \ge 10^{-2}$
marked in red). These scars control the dynamics and lead to
long-time coherent oscillation in the correlation function dynamics
(Figs.\ \ref{fig3} ({\bf a}) and ({\bf d})). At $\omega_D=8.0$
(panel {\bf b}), the scars start to merge with the thermal states
leading to thermalization in the correlator dynamics shown in Fig.\
\ref{fig3} ({\bf b}). At $\omega_D=7.75$, one finds complete absence
of scars; all states are thermal and none of them have a high
overlap with the initial state. This leads to rapid thermalization
of the correlator as shown in Fig.\ \ref{fig3} ({\bf c}).  For all
plots $\lambda=15$, $L=26$, $|\psi_0= (|\mathbb{Z}_2\rangle+
|{\bar{\mathbb{Z}}}_2 \rangle)/\sqrt{2}$, and all energies
(frequencies) are scaled in units of $w/\sqrt{2} ~(w/(\hbar
\sqrt{2}))$.} \label{fig4} \end{figure}

The result of evolution of $O_{22}$ in this subspace is shown in
Fig.\ \ref{fig4} near the first reentrant transition. Fig.\
\ref{fig4} ({\bf a}) shows non-ergodic persistent oscillatory
dynamics at $\omega_D=8.25$. As we reduce $\omega_D$, these
oscillations dampen (Fig.\ \ref{fig3} ({\bf b})); such a behavior
can be interpreted as a precursor to ergodic dynamics and
thermalization. Upon further reduction of $\omega_D$, ergodic
dynamics consistent with ETH sets in and the fastest thermalization
is seen around $\omega_D=7.75$ (Fig.\ \ref{fig3} ({\bf c})).
Finally, at lower $\omega_D$, the persistent oscillations return
(Fig.\ \ref{fig3} ({\bf d})).

Such transitions between non-ergodic and ergodic dynamics as a
function of drive frequency can be tied to the presence or absence
of scars in the spectrum of $H_F$. This is shown in Fig.\
\ref{fig4}. Figs.\ \ref{fig4} ({\bf a})($\omega_D=8.25$) and ({\bf
d}) ($\omega_D=7.5$) clearly indicate the presence of scars having
high overlap with $(|\mathbb{Z}_2\rangle+ |{\bar{\mathbb{Z}}}_2
\rangle)/\sqrt{2}$. This is consistent with the presence of
non-ergodic dynamics characterized by persistent long-time
oscillations (Figs.\ \ref{fig3} ({\bf a}, {\bf d})). These scar
states start to merge with the thermal band around $\omega_D=8.0$
(Fig.\ \ref{fig4} ({\bf b})) indicating precursor to the thermal
behavior (Fig.\ \ref{fig3} ({\bf b})). Fig.~\ref{fig4} ({\bf c}) at
$\omega_D=7.75$ shows complete absence of scars resulting in ergodic
dynamics of $O_{22}$ and fast thermalization predicted by ETH (Fig.\
\ref{fig3} ({\bf c})).

\begin{figure}
\rotatebox{0}{\includegraphics*[width=\linewidth]{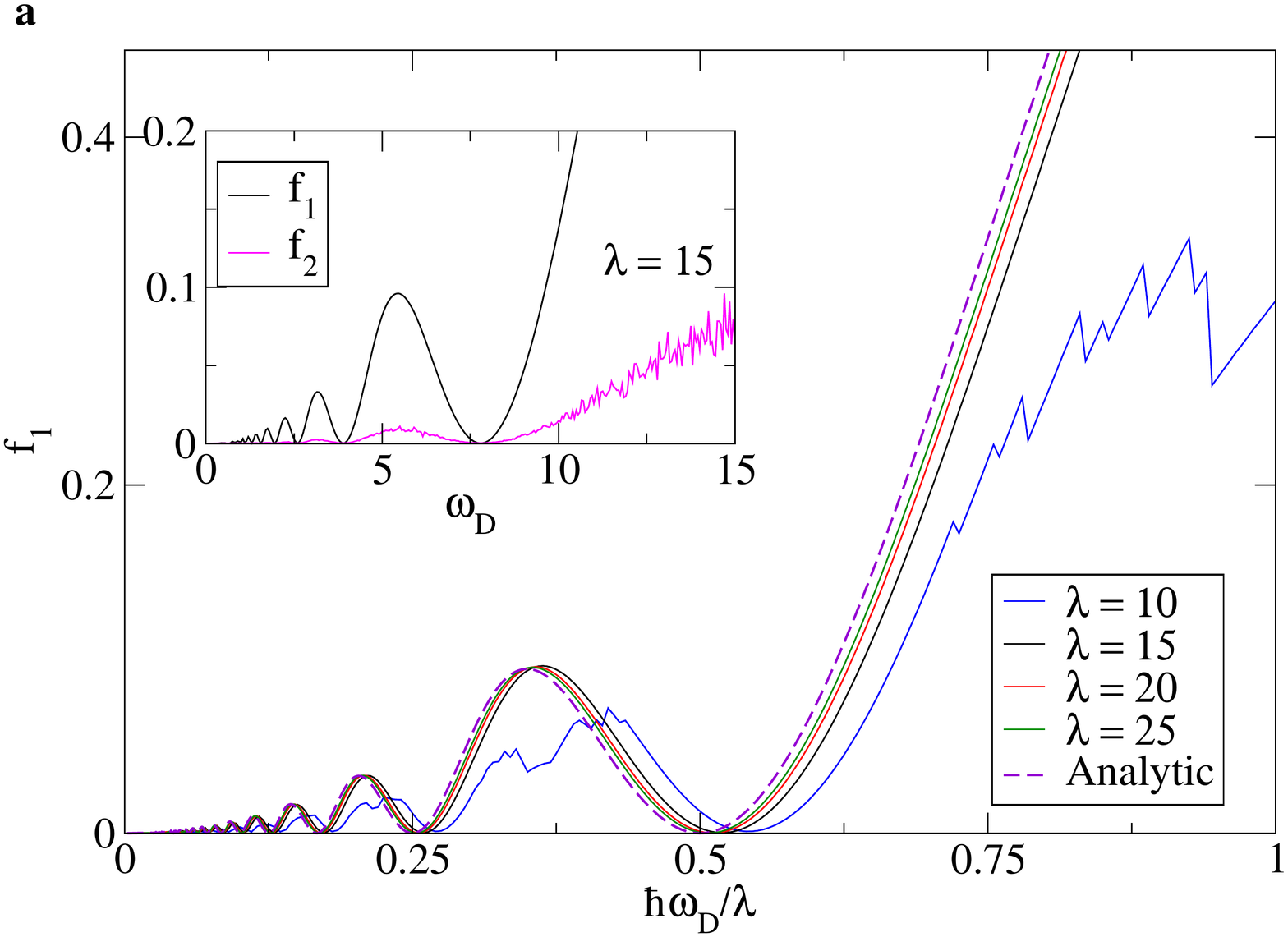}} \\
{\includegraphics*[width=0.9\linewidth]{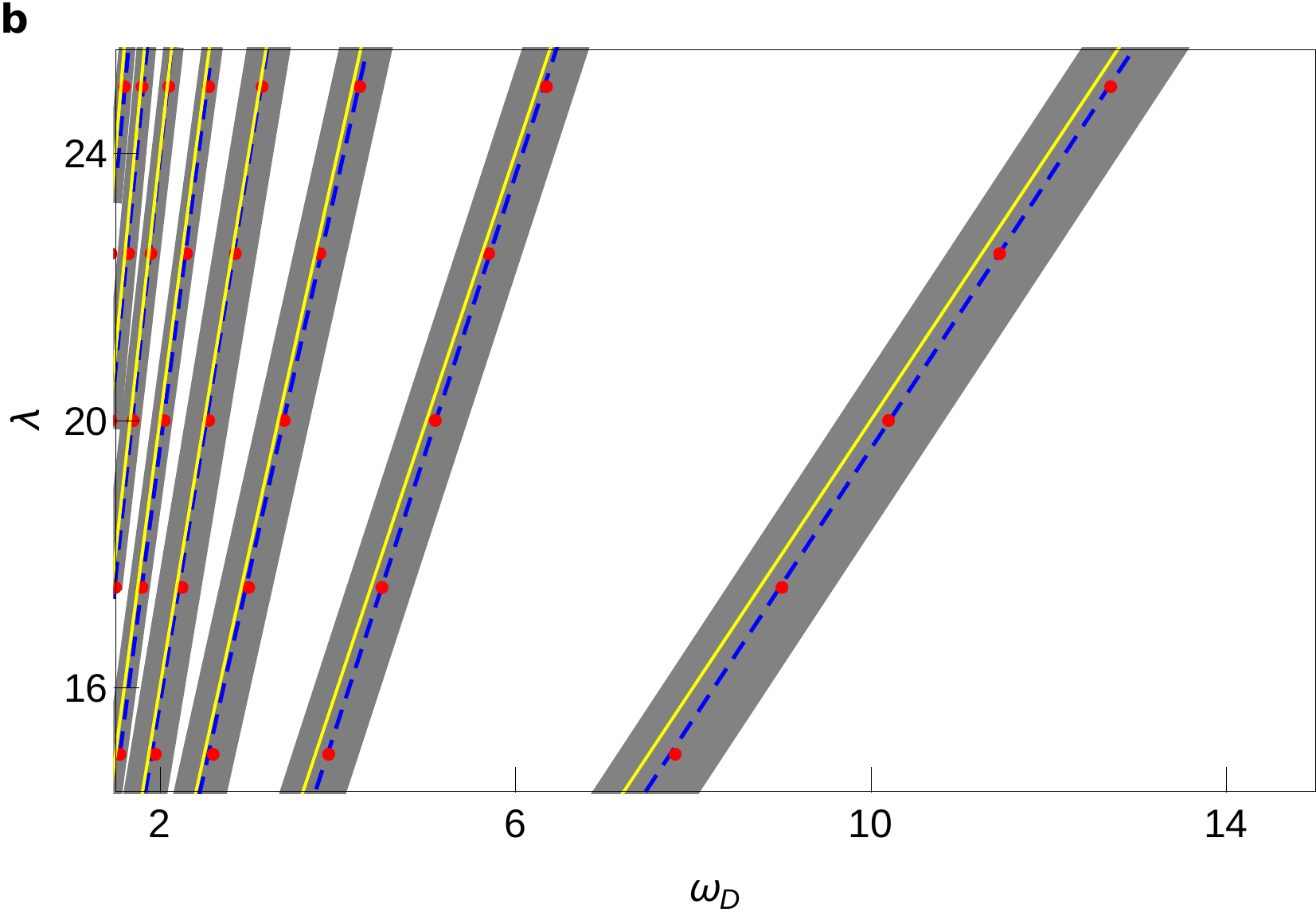}} \caption{{\bf
a}: Plot of the norm of the PXP terms in the Floquet Hamiltonian
$f_1$ as a function of $\hbar \omega_D/\lambda$ for several
$\lambda$ showing the positions of their minima. The plot
demonstrates that $f_1$ has a weak dependence on $\delta$ for
$\gamma \gg \delta$ and that $\lambda=10$ is outside this regime.
The inset shows that the norm of the PXP ($f_1$) [black solid line]
and non-PXP ($f_2$) [magenta solid line] terms become comparable
near these minima. {\bf b}: Phase diagram showing reentrant
transitions between ergodic and non-ergodic regimes as a function of
$\omega_D$ and $\lambda$. The red dots joined by dashed blue lines
indicate positions of the minima of $f_1$ ($\omega_D=\omega_D^c$)
where $O_{22}$ displays fastest thermalization, while the grey
shaded regions indicate range of $\omega_D$ around $\omega_D^c$ for
which it shows a precursor to thermalization. The yellow solid lines
indicate the positions of the transition obtained analytically from
Eq.\ \eqref{highPXP}. The white regions are where $O_{22}$ shows
long-time persistent oscillations. At low $\omega_D$, the shaded
regions cover the phase diagram almost entirely indicating absence
of coherent dynamics. All energies (frequencies) are scaled in units
of $w/\sqrt{2} ~ (w/(\hbar \sqrt{2}))$.} \label{fig5}
\end{figure}

To obtain a qualitative understanding of these transitions, we
compute the norm of the PXP-like terms in the Floquet Hamiltonian
numerically using $L=14$. To this end, we write the matrix
representation of $H_F$ in the basis states $|\phi_n\rangle$ of
$\sigma^z$ and identify the matrix elements that have $\langle
\phi_n|\sum_{\ell} \tilde{\sigma_{\ell}}^{x/y}|\phi_m\rangle \ne 0$.
Let us denote this set as ${\mathcal N}_0$ which has $N_0=
2LF_{L-1}$ elements (where $F_L$ is a Fibonacci number defined by
$F_n+F_{n+1}=F_{n+2}$ with $F_1 = F_2 = 1$). We then define $f_1$
and $f_2$ as
\begin{eqnarray} f_{1[2]} &=& \frac{1}{N_0}\sum_{\{n,m\} \in [{\not \in}]
{\mathcal N}_0} |\langle \phi_n |H_F |\phi_m \rangle|^2.
\label{normdef} \end{eqnarray}
Clearly, $f_2$ represents the contribution from the non-PXP type
of terms in $H_F$. We note that in general $f_1$ will also have
contributions from non-PXP terms since such terms may have non-zero
matrix elements for some states included in ${\mathcal N}_0$.
However, at large $\lambda/w$, the contributions from these terms are
expected to be small by at least ${\mathcal O}(w^2/\lambda^2)$.
In fact from Eq.\ \eqref{highPXP}, to leading order, we find that
\begin{eqnarray} f_1 ~=~ w^2 \frac{\sin^2 (\gamma)}{\gamma^2}, \mbox{~~~~~}
f_2 ~=~ 0. \label{f1result} \end{eqnarray} To numerically verify
that this is indeed the case, we plot $f_1$ as a function of $\hbar
\omega_D/\lambda$ for several $\lambda/w$ (Fig.\ \ref{fig5} ({\bf
a})). These curves coincide indicating that $f_1$ is almost
independent of $\delta$. Thus in this regime $f_1$ receives
negligible contributions from the non-PXP terms in $H_F$ which
necessarily depend on $\delta$ (Fig.\ \ref{fig5} ({\bf a}) also
shows that $\lambda \sim 10$ lies outside this regime).

Fig.\ \ref{fig5} ({\bf a}) and Eq.\ \eqref{f1result} also show that
$f_1$ is an oscillatory function of $\omega_D$. From the inset of
Fig.\ \ref{fig5} ({\bf a}), we find that $f_1 \simeq f_2$ near
the minima of $f_1$ at $\hbar \omega_D=\lambda/(2n_0)$ (Eq.\
\eqref{f1result}) where $n_0$ is a positive integer; in other regions,
$f_1 \gg f_2$. The ergodic dynamics of $O_{22}$ always occur in a
finite frequency interval around $\omega_D^c$ (where $f_1=f_2$).
This observation sheds light on the cause of the transitions. We
note that the PXP Hamiltonian supports scars; consequently, for
$\omega_D$ where $H_F$ is PXP-like ($f_1 \gg f_2$), one expects
the presence of scars among its eigenstates. These scars lead to
coherent non-ergodic dynamics. In contrast, near the minima of
$f_1$, where $f_1 \simeq f_2$, $H_F$ receives significant
contributions from the non-PXP terms. In the presence of such terms
which can be long-ranged at low $\omega_D$, $H_F$ does not support
scars. The bulk of its eigenstates around $\omega_D=\omega_D^c$ are
thermal; consequently the dynamics of $O_{22}$ displays ergodic behavior
consistent with the prediction of ETH. In fact, the level spacing
statistics of the eigenvalues of $H_F$ follow a Gaussian orthogonal
ensemble for $\omega_D \approx \omega_D^c$ as expected for an
ergodic system (see Supplementary Information). The behavior of the
fidelity $|\langle \Psi_n|\Psi_0 \rangle|^2$, where $|\Psi_n
\rangle$ is the state after $n$ drive cycles, across the transition
is also discussed in the Supplementary Information.

The schematic phase diagram for these reentrant ergodic to non-ergodic
transitions is summarized in Fig.\ \ref{fig5} ({\bf b}). The ergodic
regions, where $O_{22}$ displays either complete ergodic behavior or a
precursor to thermalization, are located in a small frequency window
(shown schematically in grey) around $\omega_D^c$ (indicated as red
dots joined by blue dashed lines in Fig.\ \ref{fig5} ({\bf b})). The
yellow lines indicate the positions of the transitions as obtained from
Eq.\ \eqref{highPXP} (i.e., $\lambda /\hbar \omega_D$ is an even integer),
while the white regions denote the ranges of
$\omega_D$ where $O_{22}$ shows non-ergodic oscillatory dynamics due
to the presence of scars in $H_F$. The thermal regions become denser
with decreasing $\omega_D$ and ultimately merge into a continuum at
sufficiently small $\omega_D$ where non-ergodic
coherent dynamics of $O_{22}$ ceases to exist.

{\it Discussion}: To summarize, we have studied the kinematically
constrained PXP model, a paradigmatic model for many-body
eigenstates called quantum scars that violate ETH, in the presence
of a pulsed transverse magnetic field that varies periodically in
time. In the limit of large drive amplitude of the field, the
instantaneous Hamiltonian does not have any scars but the
corresponding Floquet Hamiltonian that controls the stroboscopic
dynamics of local quantities can still host them depending on the
drive frequency. We find (a) the presence of several non-ergodic
(characterized by a coherent oscillatory behavior of density-density
correlators and scars in the Floquet Hamiltonian) and ergodic
(characterized by a thermal non-oscillatory behavior of
density-density correlators and absence of scars) regimes as a
function of the driving frequency, and (b) the possibility of tuning
the quasienergy spacing of the scars in the non-ergodic regime as a
function of the drive frequency to control the frequency of
oscillations of the correlators. Both these features are entirely
absent in the undriven PXP model and can be tested by standard
experiments using finite-size Ryberg chains and starting from an
initial state that has one Rydberg excitation at each alternate
lattice site. In particular, these reentrant transitions from
non-ergodic to ergodic regimes by tuning the drive frequency are
possibly the first example of this kind in a system without any
spatial disorder.

The mechanism for these transitions is also rather transparent in
the large drive amplitude regime as discussed here. Using a Floquet
perturbation theory, the Floquet Hamiltonian can be grouped into
PXP and non-PXP type terms, with the non-PXP terms being
suppressed by at least the inverse of the drive amplitude. The
leading PXP terms can be resummed to all orders in the drive
frequency which shows that these can diminish and become comparable
to the non-PXP terms in the neighborhood of special drive
frequencies leading to the emergence of the thermal regime. Lastly,
on a theoretical front, it would be interesting to explore the
influence of the small non-PXP terms in the non-ergodic regime to
see whether they lead to eventual thermalization on a much longer
time scale.

{\it Acknowledgments}:
The work of A.S. is partly supported through the Max Planck
Partner Group program
between the Indian Association for the Cultivation of Science (Kolkata)
and the Max Planck Institute for the Physics of Complex Systems (Dresden).
D.S. thanks DST, India for Project No. SR/S2/JCB-44/2010 for financial support.

\section{Supplementary Material}
\beginsupplement

\section {Model Hamiltonian}
\label{modelsec}

The Hamiltonian of the Bose-Hubbard model with a tilt is given by~\cite{subir1}
\begin{eqnarray} H &=& - ~w' \sum_{\langle ij\rangle} ~(b_i^{\dagger} b_j + 
{\rm H.c.}) ~-~ \sum_i ~(\mu + E_0 i) n^b_i \nonumber\\
&& +~ \sum_i ~\frac{U}{2} n^b_i (n^b_i-1), \label{SMbham1} \end{eqnarray}
where $b_i$ ($b^\dagger_i$) denotes the boson annihilation
(creation) operator on site $i$ of a 1D chain, $n^b_i= b_i^{\dagger}
b_i$ is the boson number operator, $w'$ is the hopping amplitude of
the bosons, $E_0$ denotes the magnitude of the tilt, $\mu$ is the
chemical potential, $w$ is the hopping amplitude, and $U$ is the
on-site interaction between the bosons. The tilt can be generated
either by shifting the center of the parabolic trap confining the
bosons or by applying a linearly varying Zeeman field which couples
to the spin of the bosons. The latter variation can be made time
dependent by using a magnetic field which varies periodically in
time. It is well known that the low-energy physics of these bosons
deep inside the Mott phase, whose occupation number is denoted by
$n_0$, (here we focus on the case $n_0=1$) and where $U, E_0 \gg w',
|U-E_0|$, is given by
\begin{eqnarray} H_d ~=~ - w \sum_{\ell} ~(d_{\ell} + d_{\ell}^{\dagger}) 
~+~ \lambda \sum_{\ell} ~n_{\ell}, \label{SMdham1} \end{eqnarray}
where $d_{\ell} = b_i^{\dagger} b_j/\sqrt{n_0(n_0+1)}$ denotes a
dipole annihilation operator on link $\ell$ between sites
neighboring $i$ and $j$ on a 1D lattice, $w= \sqrt{2} w'$ for
$n_0=1$, $n_{\ell} = d_{\ell}^{\dagger} d_{\ell}$ is the dipole
number operator on link $\ell$, $w$ is the amplitude of spontaneous
dipole creation or destruction, and $\lambda$ is the chemical
potential for the dipoles. This dipole model is to be supplemented
by two constraints which make it non-integrable: $n_{\ell} \le 1$
and $n_{\ell} n_{\ell +1}=0$ for all links. The phase diagram of
this model has been studied theoretically in Ref.\
\onlinecite{subir1} and has also been experimentally verified
\cite{dipoleexp1}. It is well-known that $H_d$ support a quantum
phase transition at $\lambda_c= -1.31 w$ separating a $\mathbb{Z}_2$
symmetry broken ground state ($|\mathbb{Z}_2\rangle$) for $\lambda <
\lambda_c$ and a featureless dipole vacuum ($|0\rangle$) for
$\lambda> \lambda_c$. The non-equilibrium dynamics of this model has
also been studied for quench, ramp and periodic protocols~\cite{dipoledyn}.

The dipole model described in Eq.\ \eqref{SMdham1} also serves as an
effective model for describing the low energy physics of the Rydberg
atoms. To see this we first consider the Hamiltonian of such atoms
given by \cite{scarref1,scarref1p5,scarref2,scarref3,rydramp}
\begin{eqnarray} H_{\rm RYD} &=& \sum_i ~(\Omega \sigma_i^x + \Delta n^r_i) ~+~
\sum_{ij} ~V_{ij} n^r_i n^r_j. \label{SMrydham1} \end{eqnarray}
Here $n^r_i \le 1$ denotes the number of Rydberg excitations on a
given site, $\Delta$ is the detuning parameter which can be used to
excite an atom to its Rydberg state, $V_{ij}$ denotes the interaction
between two Rydberg excitations, $\sigma_i^x = |g_i\rangle \langle
r_i|$ denotes the coupling between the ground ($|g\rangle$) and Rydberg
excited ($|r\rangle$) states, and $\Omega$ is the corresponding coupling 
strength. In experiments~\cite{scarref1}, it is possible to reach a regime
where $V_{i,i+1} \gg \Omega, \Delta \gg V_{i,i+2}$; in this case, the 
Hamiltonian the model becomes equivalent to that of
\begin{eqnarray} H'_{\rm RYD} &=& \sum_i ~(\Omega \sigma_i^x ~+~ \Delta n^r_i),
\label{SMrydham2} \end{eqnarray}
supplemented by the constraint that $n^r_i n^r_{i+1}=0$ for all
sites $i$. Clearly, this model is equivalent to Eq.~\eqref{SMdham1} with
the identification $n^r_i \to n_{\ell}$, $\Delta \to \lambda$ and
$\Omega \to - w$.

Furthermore it is also easy to see that the dipole Hamiltonian (Eq.\
\eqref{SMdham1}) is identical to the PXP model studied in Ref.\
\onlinecite{scarref1} for $\lambda=0$. The simplest way to see this
involves mapping of the dipole operators to Ising spins via the
transformation
\begin{eqnarray} \sigma_{\ell}^z &=& 2n_{\ell}-1, \quad \sigma_{\ell}^{x(y)} 
~=~ (i) ~(d_{\ell} +(-) d_{\ell}^{\dagger}), \label{SMspintrans} \end{eqnarray}
where $\sigma^\alpha$ denote the Pauli matrices for $\alpha=x,y,z$.
Moreover, the constraint of not having dipoles on adjacent links can
be implemented via a local projection operator $P_{\ell} = (1-
\sigma_{\ell}^z)/2$~\cite{scarref1p5,scarref2}. Using these, one finds the spin
Hamiltonian
\begin{eqnarray} H_{\rm spin} &=& - w \sum_{\ell} ~P_{\ell-1} \sigma^x_{\ell}
P_{\ell+1} ~+~ \frac{\lambda}{2} \sum_{\ell} ~\sigma^z_{\ell} \nonumber\\
&=& \sum_{\ell} ~\left(- w \tilde \sigma_{\ell}^x ~+~ \frac{\lambda}{2}
\sigma_{\ell}^z\right) \label{SMspinham}, \end{eqnarray}
where $\tilde{\sigma}_{\ell}^{\alpha} = P_{\ell -1}
\sigma_{\ell}^{\alpha} P_{\ell +1}$ for $\alpha=x,y,z$, and we have
ignored an unimportant constant term while writing down the
expression for $H_{\rm spin}$. The physics of $H_{\rm spin}$ within
the constrained dipole Hilbert space is identical to that of $H_d$
and $H_{\rm RYD}$. At $\lambda=0$, $H_{\rm spin}$ reduced to the PXP
model studied in Ref.\ \onlinecite{scarref1}. Note that both the
constraints of the dipole model are incorporated in $H_{\rm spin}$
via the local projection operators $P_{\ell}$.

Eq.\ \eqref{SMspinham} has been used for the analysis in the main text.

\section{Magnus expansion}
\label{magnussec}

We consider the Hamiltonian $H_{\rm spin}$ given
by Eq.\ \eqref{SMspinham} in the presence of a periodic drive
characterized by a square pulse protocol with time period $T=
2\pi/\omega_D$, where $\omega_D$ is the drive frequency:
$\lambda(T)= -(+) \lambda$ for $t\le (>) T/2$. In what follows, we
will chart out the details of the computation of the Floquet
Hamiltonian of such a driven system using a high-frequency Magnus expansion.

For this protocol, the unitary matrix governing the evolution by a
time period is given by
\begin{eqnarray} U=e^{-iH_+T/(2\hbar)} e^{-iH_{-}T/(2\hbar)} = e^{X_+} e^{X_-} 
= U_+ U_-, \label{unitevolmag} \end{eqnarray}
where $X_{\pm}=(-iT /2\hbar) H_{\pm}$. For future reference, we
also define $X_{1,2}$ given by
\begin{eqnarray} X_{1[2]} &=& \left(\frac{i \hbar}{2T} \right) w
\left[-\frac{\lambda}{2}\right] ~\sum_{\ell} ~\tilde \sigma_{\ell}^x
[\sigma_{\ell}^z], \label{x1x2eq} \end{eqnarray}
such that $X_{\pm}= X_1 \pm X_2$. The Floquet Hamiltonian can be
obtained from $U$ as $H_{F}=(i \hbar/T) \ln U$. Using the
Baker-Campbell-Hausdorff formula, one can express
\begin{eqnarray} \ln[e^{X_+}e^{X_-}] &=&
X_+ + X_- +\frac{1}{2}[X_+,X_-]\nonumber\\
&& +\frac{1}{12}[X_+ -X_-,[X_+,X_-]]\nonumber\\
&& -\frac{1}{24}[X_-,[X_+,[X_+,X_-]]]+ \cdots. \label{BCH} \end{eqnarray}

From Eq.\ \eqref{BCH} we can find terms of different order in the Floquet
Hamiltonian. The computation of these terms up to $O(1/\omega_D^2)$
is straightforward and yields
\begin{eqnarray} H_{F}^{0}&=& - w \sum_{\ell} ~\tilde{\sigma_{\ell}^x}, \quad
H_{F}^{1} ~=~ -w \gamma \sum_{\ell} ~\tilde{\sigma_{\ell}^y} \nonumber\\
H_{F}^{2} &=& \frac{2w}{3} \gamma^2 \sum_{\ell} ~\tilde{\sigma_{\ell}^x}, 
\label{ho1} \end{eqnarray}
where $\gamma= \lambda T/(4 \hbar)$. Note that these terms lead to
a renormalized PXP model; it amounts to a change of magnitude of the
coefficient $w$ of the standard PXP Hamiltonian and also a rotation
in spin-space which depends on $\omega_D$. Note that the second
order term in the expansion has an opposite sign compared to the
zeroth order term. As we will see later, this is a general
feature of the model; any term $\sim \gamma^n$ in the renormalized
PXP model always comes with a opposite sign compared to a term
$\sim \gamma^{n-2}$.

The first non-trivial longer-ranged terms in $H_F$ arises in
$\mathcal{O}(1/\omega_D^3)$. Its derivation involves some subtle issues. To see
this, let us consider the commutator $C_1 =[X_+,[X_+,X_-]] = C_{1a} +
C_{1b}$. It is easy to see after a straightforward calculation that
\begin{eqnarray} C_{1a} &=& -2 w \lambda^2 \left( \frac{i T}{2 \hbar}\right)^3
\sum_{\ell} ~{\tilde \sigma_{\ell}^x}, \nonumber\\
C_{1b} &=& -2 w^2 \lambda \left(\frac{i T}{2 \hbar}\right)^3 \sum_{\ell} 
\left[ P_{\ell-2} \sigma_{\ell-1}^x \sigma_{\ell}^x P_{\ell} P_{\ell+1} 
\right. \nonumber\\
&& \left. + ~P_{\ell-2} P_{\ell_1} \sigma_{\ell-1}^y \sigma_{\ell}^y P_{\ell+1} 
+2 {\tilde \sigma_z^{\ell}} \right. \label{thirdexp1}\\
&& + \left. P_{\ell-1} \sigma_{\ell}^x P_{\ell} \sigma_{\ell+1}^x P_{\ell+2} 
+ P_{\ell-1} \sigma_{\ell-1}^y P_{\ell} \sigma_{\ell+1}^x P_{\ell+2} \right]. 
\nonumber \end{eqnarray}
We note that within the constrained Hilbert space, any term with
$\sigma_{\ell-1}^{\alpha} \sigma_{\ell}^{\beta} P_{\ell}$
identically vanishes for $\alpha, \beta=x,y$. Furthermore, the
projection operators $P_{\ell}$ on any link satisfy $(1-P_{\ell\pm
1}) \sigma_{\ell}^{\alpha}=0$ for $\alpha=x,y$. Using these results,
we can simplify $C_{1b}$ to obtain
\begin{eqnarray} C_{1b} &=& -2 w^2 \lambda \left(\frac{i T}{2 \hbar}\right)^3
\sum_{\ell} \left( 2 {\tilde \sigma_z^{\ell}} + {\tilde \sigma_x^{\ell}} 
{\tilde \sigma_x^{\ell}} + {\tilde \sigma_y^{\ell}} {\tilde \sigma_y^{\ell}} 
\right). \label{thirdexp2} \end{eqnarray}
Using Eq.\ \eqref{thirdexp2} and evaluating the necessary commutators, we 
finally get $H_{F}^{3} = H_{F3}^{(1)}+H_{F3}^{(2)}+H_{F3}^{(3)}+H_{F3}^{(4)}$, 
where
\begin{eqnarray} H_{F3}^{(1)} &=& \frac{(w \gamma^3 -4 \lambda \delta^3)}{3}
\sum_{\ell }\tilde{\sigma_{\ell}^y}, \label{thirdexp3} \\
H_{F3}^{(2)} &=& - ~\frac{2 \lambda \delta^3}{3} \sum_{\ell} \left[ 
\tilde{\sigma}_{\ell-1}^y {\tilde \sigma_{\ell}^{z}}+ {\tilde
\sigma_{\ell}^z} \tilde{\sigma}_{\ell+1}^y\right], \label{thirdexp4} \\
H_{F3}^{(3)} &=& - ~\frac{\lambda \delta^3}{3} \sum_{\ell}\left[
\left(\tilde{\sigma}_{j}^y\tilde{\sigma}_{j+1}^y
+\tilde{\sigma}_{j}^x\tilde{\sigma}_{j+1}^x\right)
\tilde{\sigma}_{j+1}^y \right. \nonumber\\
&& \left.+ \tilde{\sigma}_{j-1}^{y} \left(\tilde{\sigma}_{j}^y\tilde{\sigma}_{
j+1}^y +\tilde{\sigma}_{j}^x\tilde{\sigma}_{j+1}^x\right)\right], 
\label{thirdexp5} \\
H_{F3}^{(4)} &=& \frac{2\lambda \delta^3}{3} \sum_{\ell} \left(\tilde 
\sigma_{\ell}^y \sigma_{\ell+1}^z P_{\ell+1} + \sigma^z_{\ell-1} P_{\ell-1} 
\tilde \sigma_{\ell}^y \right). \label{thirdexp6} \end{eqnarray}
Here $\delta = w T/(4 \hbar)$, and we note that $\delta/\gamma = w/\lambda
\ll 1$ in the limit of large $\lambda/w$. Thus Eqs.\ \eqref{ho1} and
\eqref{thirdexp3} yield $H_0$ in the main text while Eqs.\
(\ref{thirdexp4} - \ref{thirdexp6}) yield $H_1$. This completes our
derivation of the Magnus expansion to ${\mathcal O}(1/\omega_D^3)$.

Before ending this section we note that if we concentrate on the
large $\lambda/w$ limit, it is possible to compute higher-order corrections 
to the coefficients in $H_F^0$ and $H_F^1$. This can be seen by noting
that at each order the contribution to such terms comes from
$[X_2,[X_2,[X_2, \cdots [X_2,X_1]]] \cdots]$, i.e., the $n$-th
order contribution involves commutator of $n-1$ terms involving
$\sigma_j^z$ and one $\sigma_j^x$. These commutators provide the
leading contribution in the large $\lambda/w$ limit. This structure
allows us to compute leading higher-order terms in the Magnus
expansion which contribute to the coefficients of the PXP term. A
straightforward but cumbersome computation yields
\begin{widetext}
\begin{eqnarray}
H_2 &=& -w \left( \left[ 1- \frac{2 \gamma^2}{3} + \frac{2\gamma^4}{15} - 
\frac{4 \gamma^6}{315} + \frac{2 \gamma^8}{2835} - \frac{4 \gamma^{10}}{155925}
+ \cdots \right] ~\sum_{l} ~{\tilde \sigma_{l}^x } \right. \nonumber\\
&& \left. + \gamma \left[ 1- \frac{\gamma^2}{3} + \frac{2 \gamma^4}{45} - 
\frac{\gamma^6}{315} + \frac{2\gamma^8}{14175} -\frac{2 \gamma^{10}}{467775} 
+ \cdots \right] ~\sum_{l} ~\tilde{\sigma}_{l}^{y} \right) \label{highor}.
\end{eqnarray}
\end{widetext}

It will be shown in Sec.\ \ref{fpert} that the coefficients in $H_2$
can be resummed to yield a closed form valid for arbitrary
$\omega_D$: $H_F = - (w \sin \gamma/\gamma) \sum_{j} (\cos \gamma
~\tilde \sigma^x_{j} + \sin \gamma ~\tilde \sigma^y_{j})$.

\section{Floquet perturbation theory}
\label{fpert}

We will now present a perturbation theory for a periodically
driven system~\cite{soori}. We consider a Hamiltonian $H(t) = H_0 (t) + V$,
where $H_0 (t)$ varies in time with a period $T = 2\pi/\omega$, and $V$ is a
small time-independent perturbation.
We will assume that $H_0 (t)$ commutes with itself at different times, and
will work in the basis of eigenstates of $H_0 (t)$ which are time-independent
and will be denoted as $| n \rangle$, so that $H_0 (t) | n \rangle = E_n (t)
| n \rangle$, and $\langle m | n \rangle = \delta_{mn}$. We will also
assume that $V$ is completely off-diagonal in this basis, namely, $\langle n |
V | n \rangle = 0$ for all $n$. We will now find solutions of the Schr\"odinger
equation
\beq i \hbar \frac{\partial | n (t) \rangle}{\partial t} ~=~ H(t) | n (t)
\rangle \label{sch1} \eeq
which satisfy
\beq | n (T) \rangle ~=~ e^{-i \theta_n} ~| n (0) \rangle. \label{floeig1} \eeq

For $V=0$, we have
$| n (t) \rangle = e^{-(i/\hbar) \int_0^t dt' E_n (t')} |n \rangle$,
so that the eigenvalue of the Floquet operator $U$ is given by
\beq e^{-i \theta_n} ~=~ e^{-(i/\hbar) \int_0^T dt E_n (t)}. \label{floeig2}
\eeq
We will now develop a perturbation theory to first order in $V$.
We assume that the $n$-th eigenstate can be written as
\beq |n (t) \rangle ~=~ \sum_m ~c_m (t) ~e^{-(i/\hbar) \int_0^t dt' E_m (t')}
~|m \rangle, \eeq
where $c_n (t) \simeq 1$ for all $t$, while $c_m (t)$ is of order $V$
for all $m \ne n$ and all $t$. Eq.~\eqref{sch1} implies that
\bea && i \hbar ~\sum_m \dot{c}_m (t) e^{-(i/\hbar) \int_0^t dt' E_m (t')} ~
|m \rangle \nonumber \\
& & =~ V ~\sum_m ~c_m (t) ~e^{- (i/\hbar) \int_0^t dt' E_m (t')} ~|m \rangle,
\label{sch2} \eea
where the dot over $c_m$ denotes $d/dt$. Taking the inner product of
Eq.~\eqref{sch2} with $\langle n |$ and using $\langle n | V | n \rangle = 0$,
we find that $\dot{c}_n = 0$. We can therefore choose $c_n (t) ~=~ 1$ for all
$t$. We thus have
\bea | n (t) \rangle &=& e^{-i \int_0^t dt' E_n (t')} ~|n \rangle \nonumber \\
&& +~ \sum_{m \ne n} ~c_m (t) ~e^{-i \int_0^t dt' E_m (t')} ~|m \rangle.
\label{psi1} \eea
Hence Eq.~\eqref{floeig1} implies that the Floquet eigenvalue is still given
by Eq.~\eqref{floeig2} up to first order in $V$.

Next, taking the inner product of Eq.~\eqref{sch2} with $\langle m |$, where
$m \ne n$, and integrating from $t=0$ to $T$, we get
\beq c_m (T) = c_m (0) - \frac{i}{\hbar} \langle m | V | n \rangle
\int_0^T dt e^{i \int_0^t dt' [E_m (t') - E_n (t')]}. \label{cmt1} \eeq
Since we know that Eq.~\eqref{psi1} satisfies
\beq | n (T) \rangle ~=~ e^{-(i/\hbar) \int_0^T dt E_n (t)} ~| n (0) \rangle,
\eeq
we must have
\beq c_m (T) ~=~ e^{(i/\hbar) \int_0^T dt [E_m (t) - E_n (t)]} ~c_m (0)
\label{cmt2} \eeq
for all $m \ne n$. Eqs.~(\ref{cmt1}-\ref{cmt2}) imply that we must choose
\beq c_m (0) = - \frac{i}{\hbar} \langle m | V | n \rangle \frac{\int_0^T dt
e^{(i/\hbar) \int_0^t dt' [E_m (t') - E_n (t')]}}{e^{(i/\hbar) \int_0^T dt
[E_m (t) - E_n (t)]} ~-~ 1}. \label{cmt3} \eeq
We see that $c_m (t)$ is indeed of order $V$ provided that the denominator on
the right hand side of Eq.~\eqref{cmt3} does not vanish; we will call this case
non-degenerate.

The above analysis breaks down if
\beq e^{(i/\hbar) \int_0^T dt [E_m (t) - E_n (t)]} ~=~ 1, \label{res1} \eeq
for a pair of states $|m \rangle$ and $|n \rangle$. We then have to develop
a degenerate perturbation theory. Suppose that there are $p$ states $|m
\rangle$ ($m=1,2,\cdots,p$) which have energy eigenvalues $E_m (t)$ satisfying
Eq.~\eqref{res1} for every pair of states. Ignoring all the other states of
the system for the moment, we will assume that a solution of the Schr\"odinger
equation is given by
\beq | \psi (t) \rangle ~=~ \sum_{m=1}^p ~c_m (t) ~e^{-(i/\hbar) \int_0^t dt'
E_m (t')} ~|m \rangle, \label{psit} \eeq
where we now allow all the $c_m (t)$'s to be order 1. Then we again obtain an
equation like Eq.~\eqref{sch2} except that the sum over $m$ only goes over $p$
states. To first order in $V$, we can replace $c_m (t)$ by the time-independent
constants $c_m (0)$ on the right hand side of Eq.~\eqref{sch2}. Upon
integrating from $t=0$ to $T$, this gives
\bea c_m (T) &=& c_m (0) ~-~ \frac{i}{\hbar} ~\sum_{n=1}^p ~\langle m | V | n
\rangle \nonumber \\
&& \times \int_0^T dt e^{(i/\hbar) \int_0^t dt' [E_m(t') - E_n(t')]} c_n (0).
\label{cmt4} \eea
This can be written as a matrix equation
\beq c(T) ~=~ [I ~-~ i M] ~c(0), \label{ct} \eeq
where $c(t)$ denotes the column $(c_1(t),c_2(t),\cdots,c_p(t))^T$ (where the
superscript $T$ denotes transpose), $I$ is the $p$-dimensional identity matrix,
and $M$ is a $p$-dimensional Hermitian matrix with matrix elements given by
\beq M_{mn} ~=~ \frac{\langle m | V | n \rangle}{\hbar} ~\int_0^T dt ~
e^{(i/\hbar) \int_0^t dt' [E_m(t') - E_n(t')]}. \label{mmn} \eeq
Let the eigenvalues of $M$ be given by $\mu_n$ ($n=1,2,\cdots,p$).
To first order in $V$, $I - i M$ is a unitary matrix and will therefore have
eigenvalues of the form $e^{-i\mu_n}$; the corresponding eigenstates satisfy
$c (T) = e^{-i\mu_n} ~c (0)$. Next, we want the wave function in
Eq.~\eqref{psit} to satisfy $| \psi (T) \rangle = e^{-i\theta_n } | \psi (0)
\rangle$. This implies that the Floquet eigenvalues are related to the
eigenvalues of $M$ as
\beq e^{-i\theta_n} ~=~ e^{-i\mu_n ~-~ (i/\hbar) \int_0^T dt E_n (t)}. \eeq

Given a Floquet operator $U$, we can define a Floquet Hamiltonian $H_F$ as
$U = e^{-i H_F T/\hbar}$.
Comparing this with Eqs.~\eqref{ct} and \eqref{mmn}, we see that the
matrix elements of $H_F$ are given by
\bea && (H_F)_{mn} ~=~ \frac{M_{mn}}{T} \nonumber \\
&& =~ \frac{\langle m | V | n \rangle}{T} ~\int_0^T dt ~e^{(i/\hbar)
\int_0^t dt' [E_m(t') - E_n(t')]}. \label{hmn1} \eea

We will now apply the above analysis to our model, where
\bea H_0 (t) &=& \frac{\lambda (t)}{2} ~\sum_l ~\sigma^z_l, \nonumber \\
V &=& - ~w ~\sum_l ~P_{l-1} \sigma_l^x P_{l+1}, \label{ham2} \eea
with $\lambda (t) = - \lambda$ for $0 < t < T/2$ and $+ \lambda$ for $T/2 <
t < T$. We will do Floquet perturbation theory assuming that $w \ll \lambda$.
To do this, we consider states in the $\sigma_l^z$ basis. According to the
Hamiltonian $H_0 (t)$ in Eq.~\eqref{ham2}, all such states
$| n \rangle$ have an instantaneous energy eigenvalue $E_n (t) = (\lambda (t)
/2) \sum_l \sigma^z_l$, which implies that the $\int_0^T dt E_n (t) = 0$. Thus
the unperturbed Floquet eigenvalue $e^{-i\theta_n}$ is equal to 1 for all
states; we therefore have to do degenerate perturbation theory.

If $| m \rangle$ and $| n \rangle$ are two states which are connected by the
perturbation $V$ in Eq.~\eqref{ham2}, they differ by the value of $\sigma_l^z$
at only one site and therefore $E_m (t) - E_n (t) = \lambda (t)$, assuming that
$|m \rangle$ and $| n \rangle$ have spin-up and spin-down respectively at
that site.
We find that the integral in Eq.~\eqref{cmt4} is given by
\beq \int_0^T dt ~e^{(i/\hbar) \int_0^t dt' [E_m(t') - E_n(t')]} ~=~
\frac{i2}{\lambda} ~(e^{-i\lambda T/ 2\hbar} ~-~ 1). \label{cond1} \eeq
We therefore see that if
\beq \frac{\lambda}{\hbar \omega} ~=~ 2q, \label{cond2} \eeq
where $q$ is an integer, then the expression in Eq.~\eqref{cond1} vanishes.
This means that even in degenerate perturbation theory, there is no change
in the Floquet eigenvalues and they remain equal to 1.

We will now use Eqs.~\eqref{hmn1} and \eqref{cond1}. If $|m \rangle$ and
$| n \rangle$ are two states which are connected by the perturbation $V$, we
have $\langle m | V | n \rangle = - w$. We then obtain
\bea (H_F)_{mn} &=& -~ \frac{i2 \hbar w}{\lambda T} ~(e^{-i\lambda T/2\hbar} ~
-~ 1) \nonumber \\
&=& - ~\frac{w}{\gamma} ~e^{-i \gamma} ~\sin \gamma, \label{hmn2} \eea
where $\gamma = \pi \lambda /(2 \hbar \omega)$. The Floquet Hamiltonian is
therefore given by
\beq H_F ~=~ - ~w ~\frac{\sin \gamma}{\gamma} ~\sum_n ~[ \cos \gamma ~
{\tilde \sigma}_n^x ~+~ \sin \gamma ~ {\tilde \sigma}_n^y]. \label{floham} \eeq
This vanishes if Eq.~\eqref{cond2} is satisfied; we will then have to
go to higher order in perturbation theory.

We note that Eq.\ \eqref{floham} can also be obtained by a
straightforward expansion of the evolution operators $U_{\pm}$. To
see this, we first note that for any two different sites $j$ and $j'$ we have
\begin{eqnarray} [-w {\tilde \sigma}^x_j \pm \frac{\lambda}{2} \sigma^z_j, -w {\tilde
\sigma}^x_{j'} \pm \frac{\lambda}{2} \sigma^z_{j'}] &\sim & {\mathcal O}(w^2). 
\end{eqnarray}
Thus as long as we are interested in terms of ${\mathcal O}(w)$, we can write
\begin{eqnarray} U_{\pm} &\simeq& \prod_j e^{-i T (-w {\tilde \sigma}^x_j \pm
\frac{\lambda}{2} \sigma^z_j)/2\hbar}. \label{ex1} \end{eqnarray}
One can then carry out a straightforward expansion of $U_{\pm}$. A few lines 
of algebra, required to gather terms of ${\mathcal O}(w/\lambda)$, yield
\begin{eqnarray} U_{\pm} &\simeq & \prod_j (\cos(\gamma) \mp i \sigma_z^j
\sin(\gamma) - i \frac{2w}{\lambda} \sin(\gamma) \tilde \sigma_j^x).
\label{ex2} \end{eqnarray}
Using Eq.\ \eqref{ex2}, one can compute $U=U_+ U_-$. A 
re-exponentiation of terms of ${\mathcal O}(w/\lambda)$ then yields Eq.\
\eqref{floham} in a straightforward manner.

Next, we note that the magnitude of the right hand side of Eq.~\eqref{hmn2} 
is independent of $| m \rangle$ and $| n \rangle$.
Hence
\beq \sum_{mn} ~[(H_F)_{mn}]^2 ~=~ \bigl( \sum_{mn} ~1 \bigr)~
w^2 ~\frac{\sin^2 (\gamma)}{\gamma^2}, \label{f11} \eeq
where the sum runs over all all pairs of states $m, n \in {\cal N}_0$ which
are connected by $V$.

Given a system of size $L$ and periodic boundary condition ($PBC$), and the
constraint that two up-spins cannot be on neighboring sites in any state, we
can find the number of states and the value of $\sum_{mn} 1$ in Eq.~\eqref{f11}.
To this end, we first define the Fibonacci numbers which satisfy $F_n +
F_{n+1} = F_{n+2}$, with $F_1 = F_2 = 1$; as $n$ increases, $F_n$
quickly approaches $\tau^n /\sqrt{5}$, where $\tau = (\sqrt{5}+1)/2$ is
the golden ratio. Keeping the up-spin constraint in mind, we define the
transfer matrix
\beq A ~=~ \left( \begin{array}{cc}
0 & 1 \\
1 & 1 \end{array} \right), \label{mata} \eeq
where the indices $(ij)$ of $A_{ij}$ can take values 1 (spin-up) and 2
(spin-down). The number of states in an $L$-site system is then given by
${\rm Tr} (A^L) = F_{L-1} + F_{L+1}$.
To calculate $\sum_{mn} 1$, we note that a spin at, say, site 2, can
flip between up and down only if the spins at sites 1 and 3 are both down.
The contribution of all such states to $\sum_{mn} 1$ is 2 times the number of 
all possible states for sites 4 to $L$ with open boundary condition ($OBC$) and
the up-spin constraint; the factor of 2 is because the spin at site 2 can be
up or down, giving two states. Since the number of bonds for an $OBC$ 
system with sites 4 to $L$ is $L-4$, the number of states for such
a system is given by the sum of all the matrix elements of $A^{L-4}$. This is
equal to $F_{L-1}$. In the above argument, we assumed that there the spin which
can flip between up and down is at site 2. However, the site 2 could have been
anywhere else in the $L$-site system. We therefore see that $N_0 \equiv
\sum_{mn} 1 = 2 L F_{L-1}$. This leads us to define the normalized quantity
\beq f_1 ~=~ \frac{1}{N_0} ~\sum_{mn} ~[(H_F)_{mn}]^2 ~=~ w^2 ~\frac{\sin^2
(\gamma)}{\gamma^2}, \label{f1} \eeq

\section{Symmetry of Floquet operator and zero modes}
\label{zeromode}

We will now discuss an exact symmetry of the Floquet operator
for our driving protocal. We will then see that this symmetry implies
that there will be a large number of states with zero quasienergy.

We define an operator
\beq Q ~=~ \prod_{n=1}^L ~\sigma_n^z, \eeq
which is unitary and satisfies $Q^{-1} = Q$. The eigenvalues
of $Q$ are $\pm 1$, and the corresponding eigenstates have an even (odd)
number of down spins. Next, we see that $Q$
anticommutes with the first term and commutes with the second term in
Eq.~\eqref{SMspinham}. As a result, the Floquet operator defined in
Eq.~\eqref{unitevolmag} satisfies
\beq U^{-1} ~=~ Q ~U ~Q. \label{symu} \eeq
This means that if $| \psi_n \rangle$ is an eigenstate of $U$ with eigenvalue
$e^{-i \theta_n}$, $Q | \psi_n \rangle$ will be an eigenstate of $U$
with eigenvalue $e^{i \theta_n}$. Hence, all the quasienergies must come in
$\pm$ pairs, except for quasienergies 0 and $\pi$ which correspond to Floquet
eigenvalues equal to $\pm 1$. We also see that eigenstates of $U$ with
eigenvalues $\pm 1$ can be simultaneously chosen to be eigenstates of
$Q$; hence they will be superpositions of states all of which have
an even or an odd number of down spins.

Given the Floquet Hamiltonian $H_F$ defined through $U=e^{-i H_F T/\hbar}$,
Eq.~\eqref{symu} implies that
\beq Q~ H_F ~Q ~=~ - ~H_F. \label{symhf} \eeq
Note that this is an exact symmetry, independent of the Magnus expansion or
Floquet perturbation theory. We will now use the arguments given in
Ref.~\onlinecite{scarref1p5} to argue that there will be a large number of
eigenstates of $H_F$ with zero eigenvalue; we will call these zero modes.
Eq.~\eqref{symhf} implies that $H_F$ can
be thought of as defining a tight-binding model of a particle moving on a
bipartite lattice, where the two sublattices correspond to eigenvalues of
$Q$ being equal to $\pm 1$. For such a tight-binding model, it is known
that a lower bound on the number of zero modes is given by the difference
of the number of states with $Q$ equal to $\pm 1$. 

Next, we use the
parity symmetry of our system corresponding to a reflection about the middle
of the bond connecting sites labeled $L/2$ and $L/2 + 1$ (we will assume
that $L$ is even). We define a parity transformation $P$ which
does this reflection. Given an arbitrary state $| \psi \rangle$, the parity
transformed state is $P | \psi \rangle$. The superpositions
$|\psi \rangle \pm P | \psi \rangle$ then give two states with 
even (odd) parity respectively. However, states of the product form
\beq | \psi \rangle ~=~ | \sigma_1 ~\sigma_2 ~\cdots ~\sigma_{L/2-1} ~
\sigma_{L/2} ~\sigma_{L/2} \sigma_{L/2-1} \cdots \sigma_2 \sigma_1 \rangle, 
\label{prodst} \eeq
where each $\sigma_n$ can be spin-up or down, clearly satisfy $P | 
\psi \rangle = | \psi \rangle$. Such states therefore lie in the even parity 
sector; further, they have eigenvalue of $Q$ equal to $+1$ since each 
value of $\sigma_i$ appears twice in Eq.~\eqref{prodst}.
Let the number of such states be $N'$ (we will calculate this number below).
Now, let $N_{PQ}$ denote the number of orthonormal states with $P =
\pm$ (denoting even/odd) and $Q = \pm$ (denoting $\pm 1$). We then
see that $N_{++} = N_{-+} + N'$ and $N_{+-} = N_{--}$. A lower bound on the
number of zero modes in the odd parity sector is $|N_{-+} - N_{--}|$, and in 
the even parity sector is $| N_{++} - N_{+-}| = |N_{-+} - N_{--} + N'|$. Since
\beq |N_{-+} - N_{--} + N'| ~+~ |N_{-+} - N_{--}| ~\ge~ N', \eeq
regardless of the values of $N_{-+}$ and $N_{--}$, we see that $N'$ gives 
a lower bound on the number of zero modes.

To calculate the number $N'$, we note that the string $(\sigma_1,\sigma_2,
\cdots, \sigma_{L/2})$ for the first $L/2$ sites in Eq.~\eqref{prodst} must
begin and end with a down spin to ensure that the neighboring sites
$(L/2,L/2+1)$ and $(L,1)$ do not both have spin-up. The number of of such
strings is given by the $22$ matrix element of $A^{L/2-1}$; this gives
$N' = F_{L/2}$. Thus the number of zero modes increases exponentially
with $L$, as $\tau^{L/2} /\sqrt{5}$.

\section{Half-chain entanglement}
\label{halfchainsec}

Here we detail out the procedure for computation of the
half-chain entanglement $S_{L/2}$ used in the main text. The
procedure could be applied to equilibrium or Floquet eigenstates or
to an arbitrary driven state of the given model in Eq.\ \ref{SMspinham}. 

\begin{figure}[!ht]
\includegraphics[height=80mm,width=\linewidth]{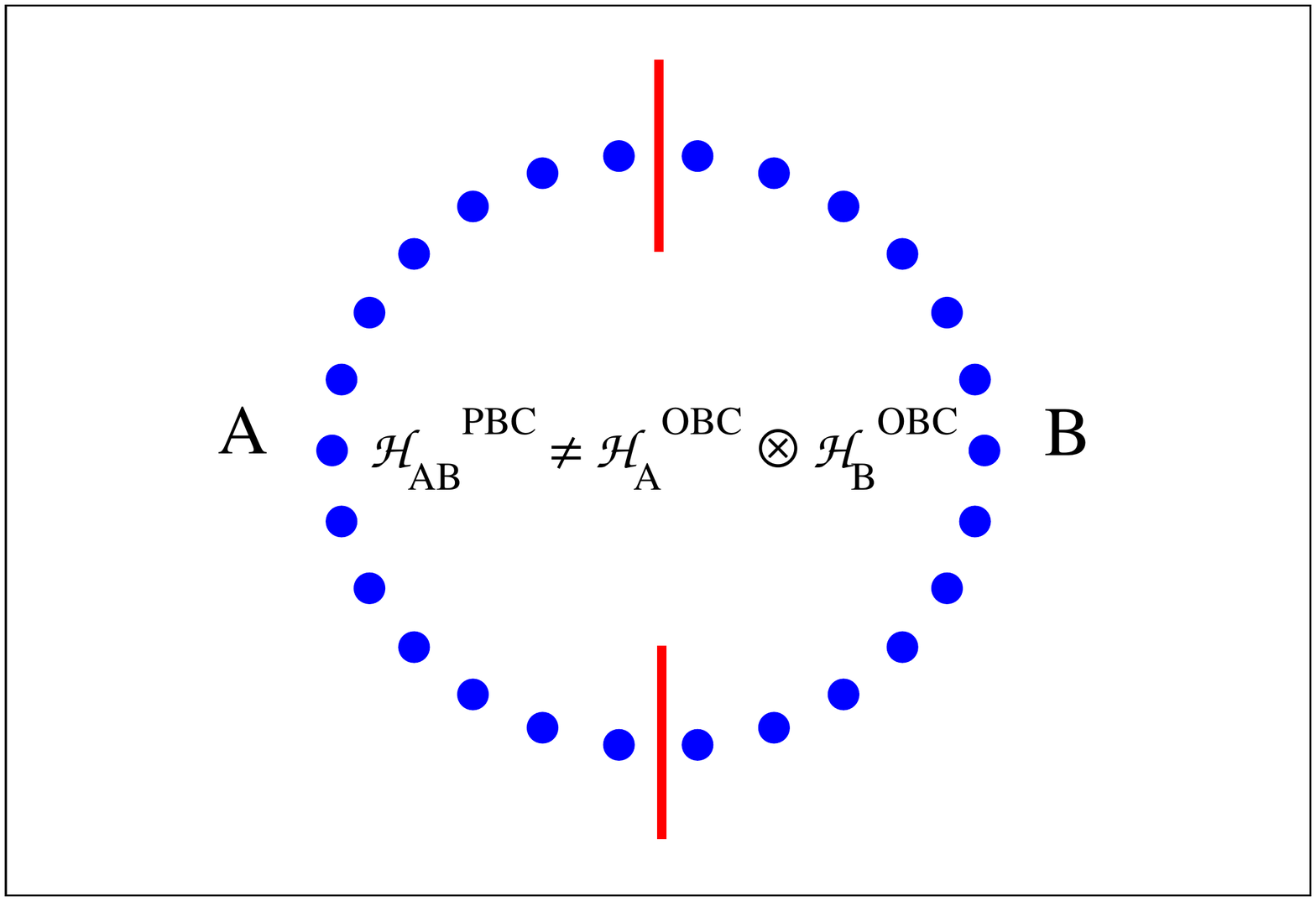}
\caption{Schematic representation of the bipartition of the one-dimensional 
model considered here (Eq.\ \ref{SMspinham}) with periodic boundary condition.}
\label{figappb1} \end{figure}

We first note that the full density matrix ($DM$) is given by
$\rho_{AB}=\ket{\psi}\bra{\psi}$ where $AB$ is the whole system of size $L$ 
with $PBC$. We divide this system into two parts $A$ and $B$ of size $L/2$ 
each with $OBC$ as schematically shown in Fig.~\ref{figappb1}.

Next, we calculate the reduced density matrix of any of the subsystems
by integrating out the other subsystem. This leads to
\begin{equation} \rho_{A(B)} ~=~ Tr_{B(A)}\rho_{AB}. \label{den1} \end{equation}
This procedure involves taking a partial trace over the environment
degrees of freedom. For example, the $(ij)$-th element,
$\rho_A(i,j)$, of the reduced density matrix is given by
\begin{equation} \rho_A(ij) ~=~ \sum_{k=1}^{HSD_B^{OBC}}\bra{i;k}\rho_{AB}
\ket{k;j}, \label{den2} \end{equation}
where $i$ and $j$ represent product states in $A$, and $k$ represent
product states in $B$ with OBC. However, since the full system ($AB$)
had PBC, the Hilbert space dimension (HSD) of $AB$ does not match the
HSD of ${\mathcal H}_A \otimes {\mathcal H}_B$ with OBC. To see
this, we note that $HSD_L^{PBC}=F_{L-1}+F_{L+1}$ and
$HSD_L^{OBC}=F_{L+2}$, where $F_L$ denotes the $L$-th
Fibonacci number. Since $F_{L+2}= F_L
+F_{L+1} > F_{L-1}+F_{L+1}$ for any $L$, one has $F_{L_1+2} F_{L_2+2}
> F_{L_1+L_2-1}+F_{L_1+L_2+1}$ for any $L_1,\,L_2$. Thus, while taking
the summation in Eq.\ \eqref{den1}, one has to exclude the states for which 
at any one of the junctions marked in red in Fig.~\ref{figappb1}, the 
end points of both $A$ and $B$ are occupied by a dipole.

Since the Hamiltonian and hence the full density matrix is in the
configuration (product) basis, any matrix element of the reduced
density matrix can be expressed as a sum of some of the elements of
full DM, and we can rewrite Eq.\ \eqref{den2} as
\begin{equation} \rho_A(ij)~=~ \sum_{k=1}^{HSD_B^{OBC}} \rho_{AB}(i;k,k;j).
\label{den3} \end{equation}
We stress here that one needs to be careful about this procedure if
the density matrix is expressed using some other basis. For example in
our case $\rho_{AB} = \sum_{m,n}\rho(m,n)\ket{m}\bra{n}$ where
$\ket{m}$,$\ket{n}$ are states in the zero total momentum ($K=0$)
and even parity ($P=+1$) sector. In this case
\begin{equation} \rho_A(ij) ~=~ \sum_{k=1}^{HSD_B^{OBC}}\bra{i;k} ~(\sum_{m,n}
\rho(m,n)\ket{m}\bra{n})~ \ket{k;j}. \label{den4} \end{equation}
This indicates that one needs to search for states
$\ket{m^{i;k}}$, $\ket{n^{j;k}}$ having non-zero overlap with the
states $\ket{i;k}$ and $\ket{j;k}$ given by
\begin{eqnarray} \ket{m^{i;k}} &=& \frac{1}{\sqrt{L_m^{i;k}}} ~(\cdots+
\ket{i;k} +\cdots), \nonumber\\
\ket{n^{j;k}} &=& \frac{1}{\sqrt{L_n^{j;k}}} ~(\cdots+\ket{j;k}+\cdots). 
\label{den5} \end{eqnarray}
In this case, $(i,j)$-th element of $\rho_A$ is given by
\begin{eqnarray} \rho_A(ij) &=& \sum_{k=1}^{HSD_B^{OBC}}
\frac{1}{\sqrt{L_m^{i;k}L_n^{j;k}}} \rho_{AB}(m^{i;k},n^{j;k}), \label{den6}
\end{eqnarray}
and the diagonalization of $\rho_{A}$ gives the Von-Neumann entropy
since $S_{A}=-\sum_{i=1}^{HSD_{A}^{OBC}}p_i \ln(p_i)$, where $p_i$ are
the eigenvalues of $\rho_A$. This procedure is used to compute
$S_{L/2}$ in the main text.

\section{Fidelity and Level statistics}
\label{fidstatsec}

\begin{figure}
\rotatebox{0}{\includegraphics*[width=\linewidth]{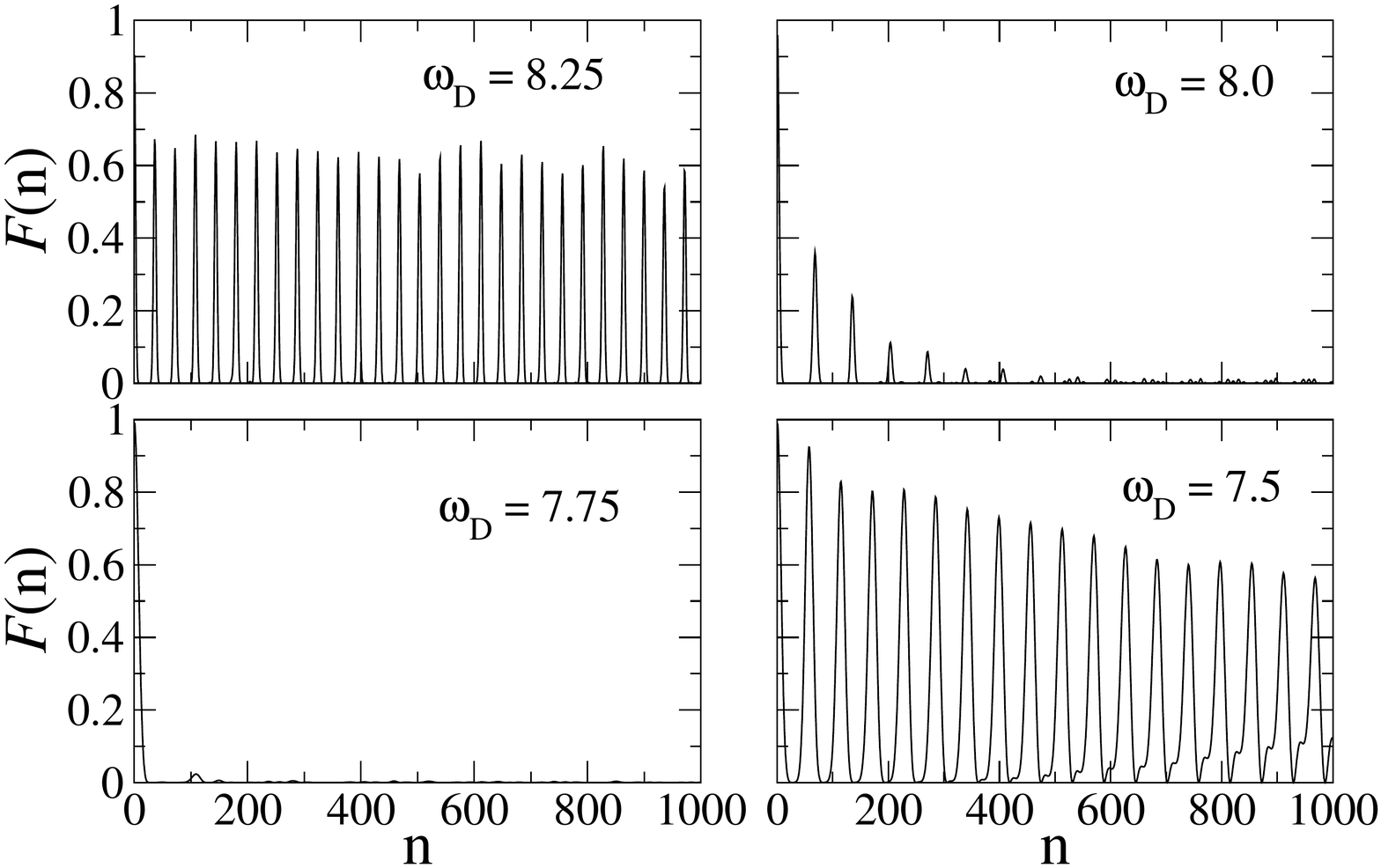}}
\caption{Plot of the fidelity ${\mathcal F}(n)$ as a function of $n$
showing periodic persistent revivals where the dynamics is
controlled by scars [top left and bottom right panels] and fast
decay with no subsequent revival in their absence [bottom left
panel]. An intermediate behavior indicating crossover from coherent
to thermal regime is shown in the top right panel. All
energies (frequencies) are scaled in units of $w/\sqrt{2}
~(w/(\hbar\sqrt{2}))$, and $L=26$, $\lambda=15$ for all plots.} \label{fig9} \end{figure}

In this section, we note that the transitions from the ergodic to
non-ergodic behaviors will also manifest themselves in the fidelity
and eigenvalue statistics.

\begin{figure}[!ht]
\includegraphics[width=\linewidth]{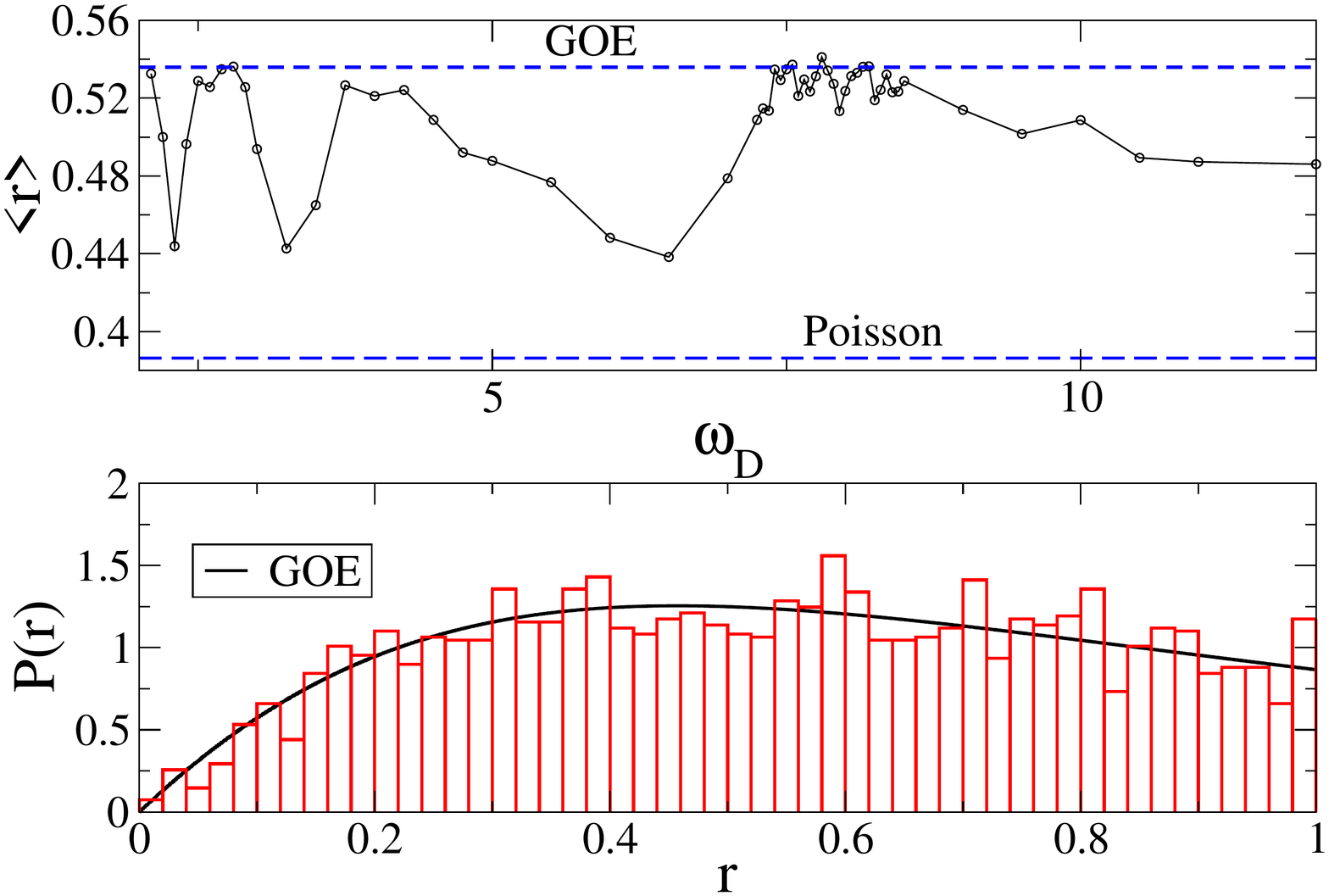}
\caption{Upper panel: A plot of $\langle r \rangle$ as computed using 
the eigenvalues of $H_F$ as a function of $\omega_D$. Lower panel: The 
distribution of $r$ at $\omega_D=7.8$ obtained using exact numerics 
(red bins) and the GOE distribution (black curve) from random matrix
theory. We have used $L=26$, $\lambda=15$, and all energies (frequencies) 
are scaled in units of $w/\sqrt{2} ~(w/(\hbar\sqrt{2}))$.} \label{rfig} \end{figure}

The fidelity of the driven Rydberg chain computed after $n$ cycles
of the drive is defined as
\begin{eqnarray} {\mathcal F}(n) ~=~ |\langle \Psi(n)|\Psi_0\rangle|^2,
\end{eqnarray}
where $|\Psi(n)\rangle$ denotes the state of the system after $n$ cycles
of the drive, and $|\psi_0\rangle$ is the initial state. In the
regime where the dynamics is dominated by scars, the presence of
long-term coherent oscillations indicates that one could expect
periodic revival of ${\mathcal F}(n)$ to values near unity; in
contrast, the thermal region, ${\mathcal F}(n)$ is expected to decay
rapidly to zero and never revive. These behaviors are numerically
confirmed in Fig.\ \ref{fig9} near the ergodic to non-ergodic
transition around $\omega_D= 7.75$. The top left ($
\omega_D = 8.25$) and the bottom right ($ \omega_D=7.5$)
panels display periodic persistent revivals of ${\mathcal F}(n)$
before and after the first transition. Such revivals are completely
absent in the bottom left panel ($\omega_D=7.75$) in the
thermal region where ${\mathcal F}(n)$ decays to zero without any
subsequent revival. The top right panel ($\omega_D=8$) shows
an intermediate behavior displaying a few (smaller) revivals and subsequent 
decay. This indicates a crossover from a coherent to a thermal regime.

Next we discuss the characteristics of eigenvalue statistics across
the transition. To this end, we first arrange the Floquet eigenvalues
($E_F^n$) (excluding the zero modes) in ascending order in the range
$[-\omega_D,\omega_D)$, and then calculate the gaps
$\delta_n=E_F^n-E_F^{n-1}$ between adjacent eigenvalues. This allows
us to compute the distribution of the ratio of successive gaps in
the energy spectrum~\cite{huse}
\begin{equation} r_n ~=~ \frac{{\rm Min}(\delta_n,\delta_{n-1})}{{\rm
 Max}(\delta_n,\delta_{n-1})}. \end{equation}
The distribution of $r_n$ for an ergodic (thermal) system obeys
the Gaussian orthogonal ensemble (GOE) and can be computed using random
matrix theory \cite{Atas} to be
\begin{equation} P_{GOE}(r) ~=~ \frac{27}{4}\frac{r+r^2}{(1+r+r^2)^{5/2}},
\end{equation}
with an average value $\langle r \rangle_{GOE}\approx0.535$. In
contrast, for a fully non-ergodic (localized) system, the
distribution of $r_n$ is Poissonian $P_{POI}(r)=2/(1+r)^2$ with
$\langle r\rangle_{POI}\approx0.386$. In Fig.\ \ref{rfig}, we plot
$\langle r \rangle$ vs $\omega_D$ computed using the eigenvalues of
$H_F$. The plot indicates that $\langle r\rangle$ reaches its GOE form
precisely at the transition points; it remains lower than this value
for other $\omega_D$. It is to be noted that $\langle r \rangle$
never reaches its Poisson value which indicates non-integrability of
the system for all $\omega_D$. This confirms the signature of the
transition in the level statistics.

\end{document}